\documentclass[aps,prl,twocolumn,groupedaddress,nofootinbib,preprintnumbers]{revtex4}
\usepackage{graphicx}
\usepackage{dcolumn}
\usepackage{bm}
\usepackage{latexsym}
\usepackage{amsfonts}
\usepackage{amssymb}
\usepackage{amsmath}
\usepackage{verbatim}
\usepackage{color}

\begin{document}

\preprint{YITP-15-25}

\title{Resonant Primordial Gravitational Waves Amplification}

\author{Chunshan Lin}
\affiliation{Yukawa Institute for Theoretical Physics, Kyoto University}
\author{Misao Sasaki}
\affiliation{Yukawa Institute for Theoretical Physics, Kyoto University}

\begin{abstract}
We propose a mechanism to evade the Lyth bound in models of inflation.
We minimally extend the conventional single-field inflation model in general 
relativity (GR) to a theory with non-vanishing graviton mass in the very 
early universe. The modification primarily affects
the tensor perturbation, while the scalar and vector perturbations are the same 
as the ones in GR with a single scalar field at least at the level of 
linear perturbation theory. During the reheating stage, the graviton mass 
oscillates coherently and leads to resonant amplification of the 
primordial tensor perturbation.
 After reheating the graviton mass vanishes and we recover GR. 
\end{abstract}

\maketitle

{\bf Introduction~} 
Inflation \cite{Guth:1980zm} is the leading paradigm of very early universe 
cosmology, but its physical origin is still mysterious. The generation of
primordial gravitational waves is a generic prediction of the inflationary 
universe. It leads to B mode polarization in the CMB, and provides an 
important window to the physics of very early universe.  
It was reported that the primordial tensor-to-scalar ratio is 
$r<0.11$(95\% CL), based on $Planck$ full sky survey \cite{Ade:2015lrj}.
 Several next-generation satellite missions (CMBPol, COrE and LiteBIRD) 
as well as the ground based 
experiments (AdvACT, CLASS, Keck/BICEP3, Simons Array, SPT-3G) 
and balloons (EBEX, Spider), are aimed at measuring primordial gravitational 
waves down to $r\sim 10^{-3}$. See Ref.~\cite{Creminelli:2015oda} for 
a recent updated forecast on these future experiments. 

According to the Lyth bound \cite{Lyth:1996im}, the tensor-to-scalar ratio 
is propotional to the variation of the inflaton field during inflation, 
i.e. $\Delta\phi/M_p\simeq \int dN\sqrt{r/8}$.
The threshold $\Delta \phi = M_p$ then corresponds to $r=2\times 10^{-3}$, 
assumed that tensor power spectrum is nearly scale-invariant.  
The sizeable amplitude of the primordial gravitational waves requires 
a super-Planckian excursion of the inflaton, i.e.  $\Delta \phi > M_p$.  
 
In quantum field theory, the naturalness principle tells us that
 the variation of a field $\phi$ over the distance greater than the cutoff 
scale is generally regarded as being out of the validity of the theory.
In a gravitational system, we take the Planck mass as the UV cutoff scale,
because gravity strongly couples to the matter sector and
the graviton-graviton scattering violates unitarity above this scale. 
Thus the inflationary prediction may not be reliable in the case of 
a super-Planckian excursion.
Therefore, the detection of the primordial tensor perturbation
with its amplitude larger than the threshold value $r=2\times 10^{-3}$
has a profound impact on our understanding of fundamental physics.
It implies that either quantum field theory or gravity may be modified 
in the very early universe.

In this letter, by means of modifying gravity, we propose a new mechanism
to evade the Lyth bound. We consider a minimal extension of GR 
with a non-vanishing graviton mass term in the very early universe. 
Specifically we propose a model in which the graviton mass is proportional
to the inflation during reheating. Then the coherent oscillation of 
the inflaton induces that of graviton mass and gives rise to
resonant amplification of the primordial tensor perturbation.
This is a broad parametric resonance which includes all long wavelength modes, 
given the graviton mass is much greater than the Hubble constant during 
reheating. After reheating, the graviton mass vanishes as the inflaton
decays and we recover GR.

{\bf A massive gravity theory~} The theoretical and observational 
consistency of massive gravity has been a longstanding problem,  
the pioneering attempt could be traced back to  Fierz and Pauli's work 
in 1939 \cite{Fierz1939}. However, Fierz-Pauli's theory and its non-linear 
completion, the so called dRGT massive gravity \cite{deRham:2010kj},
suffer from many 
pathologies \cite{vdvz1,vdvz2,bdghost,Gumrukcuoglu:2011zh,DeFelice:2012mx}.
The origin of these pathologies is probably the Poincare symmetry 
of the $St\ddot{u} ckelberg$ scalar field configuration.

Away from the Poincare symmetry, a broad class of massive gravity theories 
have been discussed in the 
literature \cite{Dubovsky:2004sg,Rubakov:2004eb,Comelli:2013txa,Comelli:2014xga,Lin:2013aha,Lin:2013sja,Lin:2015cqa}.
In this letter, we consider a massive gravity theory with the internal 
symmetry \cite{Dubovsky:2004sg}\cite{Dubovsky:2004ud}
\begin{eqnarray}\label{sym1}
\varphi^i\to\Lambda_j^i\varphi^j, \qquad
\varphi^i\to\varphi^i+\Xi^i\left(\varphi^0\right),
\end{eqnarray}
where $\Lambda_j^i$ is the $SO(3)$ rotational operator, 
$\Xi^i\left(\varphi^0\right)$ are three arbitrary functions of 
their argument, $\varphi^i$ and $\varphi^0$ are four 
$St\ddot{u} ckelberg$ scalars with non-trivial VEVs,
\begin{eqnarray}
\varphi^0=f(t),~~~~\varphi^i=x^i, ~~i=1,2,3.
\end{eqnarray} 
These nontrivial VEVs give a non-vanishing graviton mass. 
At the first derivative level, there are two combinations of 
the $St\ddot{u} ckelberg$ fields that respect this symmetry, 
\begin{eqnarray}\label{Zij}
X&=&g^{\mu\nu}\partial_{\mu}\varphi^0\partial_{\nu}\varphi^0,\nonumber\\
Z^{ij}&=&g^{\mu\nu}\partial_{\mu}\varphi^i\partial_{\nu}\varphi^j
-\frac{g^{\mu\nu}\partial_{\mu}\varphi^0\partial_{\nu}\varphi^i
\cdot g^{\lambda\rho}\partial_{\lambda}\varphi^0\partial_{\rho}\varphi^j}{X}.
\end{eqnarray}
The graviton mass term could be written as a generic scalar function of 
the above two ingredients. 

Due to the internal symmetry $\varphi^i\to\varphi^i+\Xi^i\left(\varphi^0\right)$, 
there are only 3 dynamical degrees of freedom (DOF) in our theory, i.e. 
2 tensor modes, and 1 scalar mode. In the 
language of ADM formalism or the ($3+1$)-decomposition of space-time,
we find that these two ingredients in Eq.~(\ref{Zij}) are free from the
shift $N^i$ and thus the associated Hamiltonian of gravity is linear in $N^i$. 
This implies that 3 momentum constraints and the associated secondary constraints  eliminate 3 DOF in $h_{ij}$, and the number of residual DOF
is thus 3 \cite{Comelli:2014xga}.

Now we apply this massive gravity theory to the early universe. 
To minimize our model, we identify the time-like $St\ddot{u} ckelberg$ scalar 
with the inflaton scalar field $\phi$, i.e $\varphi^0=\phi$. 
By doing this, we achieve a minimal model of massive gravity, in which
 only the tensor modes receive a modification, while the scalar and vector 
modes remain the same as the ones in the single scalar model in GR. 

To be specific, we consider the following action with enhanced global symmetry $\varphi^i\to\text{constant}\cdot\varphi^i$,
\begin{eqnarray}\label{action}
S=\int d^4x\sqrt{-g}\left[\frac{M_p^2}{2}\mathcal{R}
-\frac{1}{2}g^{\mu\nu}\partial_{\mu}\phi\partial_{\nu}\phi
-V\left(\phi\right)\right.\nonumber\\
\left.-\frac{9}{8}M_p^2m_g^2\left(\phi\right)
\frac{\bar{\delta}Z^{ij}\bar{\delta}Z^{ij}}{Z^2}\right],
\end{eqnarray}
where $V(\phi)$ is the inflaton potential, 
the numerical factor $9/8$ is inserted for later convinience, 
and $\bar{\delta}Z^{ij}$ is a traceless tensor defined by \cite{Lin:2015cqa}
\begin{eqnarray}
\bar{\delta}Z^{ij}\equiv Z^{ij}-3\frac{Z^{ik}Z^{kj}}{Z},
\end{eqnarray}
where $Z^{ij}$ is defined by Eq.~(\ref{Zij}) with $\varphi^0$ replaced by 
$\phi$, $Z\equiv Z^{ij}\delta_{ij}$, and the summation over repeated indices 
is understood. Noted that the 2nd line of Eq.~(\ref{action}) is the graviton 
mass term, which does not contribute to the background energy momentum 
tensor. Its non-trivial contribution starts from the quadratic action in
 perturbations. 

As for the mass parameter $m_g^2\left(\phi\right)$, we assume the 
following scalar field dependence:
\begin{eqnarray}\label{mass}
m_g^2(\phi)=\frac{\lambda\phi^2}{1+({\phi}/{\phi_*})^4},
\end{eqnarray}
where $\phi_*$ is the inflaton field value at the end of inflation.
Without loss of generality, we assume $\phi=0$ is the minimum of the 
potential at which the inflaton settles down after reheating.

As usual, we consider a flat FLRW background,
\begin{eqnarray}
ds^2=-dt^2+a^2d\bm{x}^2.
\end{eqnarray}
Due to the $SO(3)$ rotational symmetry of the 3-space, 
we can decompose the metric perturbation into scalar, vector, 
and tensor modes. These modes are completely decoupled at linear 
order.  We define the metric perturbation variables as
\begin{eqnarray}
g_{00}&=&-\left(1+2\alpha\right)~,\nonumber\\
g_{0i}&=&a(t)\left(S_i+\partial_i\beta\right)~,\nonumber\\
g_{ij}&=&a^2(t)\left[\delta_{ij}+2\psi\delta_{ij}
+\partial_i\partial_jE\right.\nonumber\\
&&~~~~~~~~~~\left.+\frac{1}{2}(\partial_iF_j+\partial_jF_i)
+\gamma_{ij}\right]~,
\end{eqnarray}
where $\alpha$, $\beta$, $\psi$ and $E$ are scalar, 
$S_i$ and $F_i$ are vector, and $\gamma_{ij}$ is tensor. 
The vector modes satisfy the transverse condition, 
$\partial_iS^i=\partial_iF^i=0$, and
the tensor modes satisfy the transverse and traceless condition, 
$\gamma^i_i=\partial_i\gamma^{ij}=0$.

{\bf Tensor perturbation~}  The action for the tensor perturbation reads
\begin{eqnarray}
S_T^{(2)}=\frac{M_p^2}{8}\int dt d^3x a^3
\left[\dot{\gamma}_{ij}\dot{\gamma}^{ij}
-\left(\frac{k^2}{a^2}+m_g^2\right)\gamma_{ij}\gamma^{ij}\right].
\end{eqnarray}
We see that the graviton receives a mass correction. 
We quantize the tensor mode as
\begin{eqnarray}
\gamma_{ij}(x)=\sum_{s=\pm}\int d^3k
\left[a_{\bm{k}}e_{ij}(\bm{k},s)\gamma_{k}e^{i\bm{k}\cdot\bm{x}}+h.c.\right],
\end{eqnarray}
where $a_{\bm{k}}$ is the annihilation operator
and $e_{ij}(\bm{k},s)$ is the transverse and traceless polarization tensor
 which we normalize as
\begin{eqnarray}
e_{ij}(k,s)e^{ij}(k,s')=\delta_{ss'}~.
\end{eqnarray}
The equation of motion for the tensor modes reads
\begin{eqnarray}\label{eomGW}
\ddot{\gamma}_k+3H\dot{\gamma}_k+\left(\frac{k^2}{a^2}+m_g^2\right)\gamma_k=0.
\end{eqnarray}

During inflation, the universe undergoes a superluminal expansion with 
a nearly constant Hubble parameter. The vacuum fluctuations are stretched 
and frozen on super-horizon scales.   At this stage, the graviton mass is
\begin{eqnarray}
m_g^2\simeq\lambda\phi_*^2(\phi_*/\phi_i)^{2},\quad\text{because}\quad
\phi_*^4\ll\phi_i^4,
\end{eqnarray}
where the subscript $``i"$ is for ``inflation".
If the graviton mass were greater than the Hubble parameter $H$, 
the tensor modes would decay exponentially on super-horizon scales, 
and we would never see any signals today. To avoid the exponential suppression, 
the graviton mass must be much smaller than $H$ during inflation, 
\begin{eqnarray}\label{condition1}
m_g^2\simeq\lambda\phi_*^2(\phi_*/\phi_i)^{2}\ll H_i^2.
\end{eqnarray} 
As we shall see below, this condition can be easily satisfied in our model.
In passing, we note that as the scalar and vector modes are the same as in GR,
our theory is free from the Higuchi ghost \cite{Higuchi:1986py}
 even for $m_g<2H$ in the de-sitter space-time. 

Given the small but non-vanishing mass, the inflationary tensor
 spectrum is calculated by 
\begin{eqnarray}
P_{\gamma}=\frac{2H^2}{\pi^2M_p^2}\left(\frac{k}{aH}\right)^{2m_g^2/3H^2},
\end{eqnarray}
with the tilt 
\begin{eqnarray}\label{ttilt}
n_t\simeq -2\epsilon+\frac{2m_g^2}{3H_{i}^2},
\end{eqnarray}
where $\epsilon\equiv -\dot{H}/H^2$  is the slow-roll parameter.

At the end of inflation, the slow-roll condition breaks down and
 the universe undergoes reheating.
At the reheating stage, the inflaton oscillates around the potential
minimum, and gradually decays to radiation. 
The potential is expanded around the minimum as
\begin{eqnarray}
V(\phi)\simeq \frac{1}{2}M^2\phi^2+\cdots.
\end{eqnarray}
where $M$ is the mass of the inflaton during reheating 
and the dots stand for higher order corrections which are 
irrelevant at low energy scale. 
Asymptotically for large $Mt\gg1$, we have 
\begin{eqnarray}
\phi_r\simeq \frac{\phi_{*}}{\sqrt{3\pi}Mt}\sin\left(Mt\right)
\exp\left(-\frac{1}{2}\Gamma M t\right),
\end{eqnarray}
where subscript $``r"$ stands for ``reheating". 
The Einstein equations tell us $H\simeq \frac{2}{3t}$ at this stage.
To include the effect of decaying inflaton, we have simply added a decaying 
factor $e^{-\Gamma M t}$ into the above solution, 
without specifying the detailed model of reheating.

During the reheating stage, we have 
\begin{eqnarray}
\phi_{r}\ll\phi_*, ~~~\text{and thus}~~~m_g^2\simeq\lambda\phi_{r}^2.
\end{eqnarray}
The equation of motion of gravitational waves (\ref{eomGW}) becomes 
a Mathieu-type equation,
\begin{eqnarray}\label{mathieuGW}
\frac{d^2\gamma_k}{dx^2}+\frac{2}{x}
\frac{d\gamma_k}{dx}+\frac{\xi \cdot e^{-\Gamma x} }{x^2} \sin^2(x)\gamma_k=0,
\end{eqnarray}
where $x\equiv Mt $ and $\xi\equiv \frac{\lambda \phi_{*}^2}{3\pi M^2}$.
Noted that we have neglected the spatial gradient term since we are 
interested in the long wavelength modes. 

It is well known that the Mathieu equation has a very efficient and 
broad parametric resonance if $\frac{\xi\cdot e^{-\Gamma x}}{x^2}\gg1$, 
i.e. the graviton mass must be much greater than Hubble parameter 
during reheating. Let us check whether this condition can be satisfied.
 Note that the Friedmann equation tells us $M_p^2H_{r}^2\sim M^2\phi_{r}^2$,
where $H_{r}$ is the Hubble parameter during reheating.
Generally we expect that $M^2\sim H_{i}^2$ due to the breaking of 
the slow-roll condition at the end of inflation. We thus get 
\begin{eqnarray}
m_g^2\simeq \lambda\phi_{r}^2\sim
 \lambda\cdot \frac{M_p^2}{H_{i}^2}\cdot H_{r}^2. 
\end{eqnarray}
Demanding that the graviton mass be much greater than the Hubble parameter
during reheating yields the condition,
\begin{eqnarray}\label{condition2}
 \lambda \frac{M_p^2}{H_{i}^2}\gg 1.
\end{eqnarray}
Combinning conditions (\ref{condition1}) and (\ref{condition2}), we get 
\begin{eqnarray}
\frac{H_{i}^2}{M_p^2}\ll\lambda\ll
\frac{H_{i}^2}{M_p^2}\cdot\frac{M_p^2\phi_{i}^2}{\phi_*^4}.
\end{eqnarray}
Thus  $\frac{M_p^2\phi_{i}^2}{\phi_*^4}\gg1$,  i.e. $\phi_*\ll M_p$ is 
required for the self-consistency of the above inequality. 
Note that $\phi_*\ll M_p$ is also the condition of the validity of 
our effective field theory,  which is automatically satisfied for 
many small field inflationary models.  

\begin{figure}
\begin{center}
\includegraphics[width=0.25\textwidth]{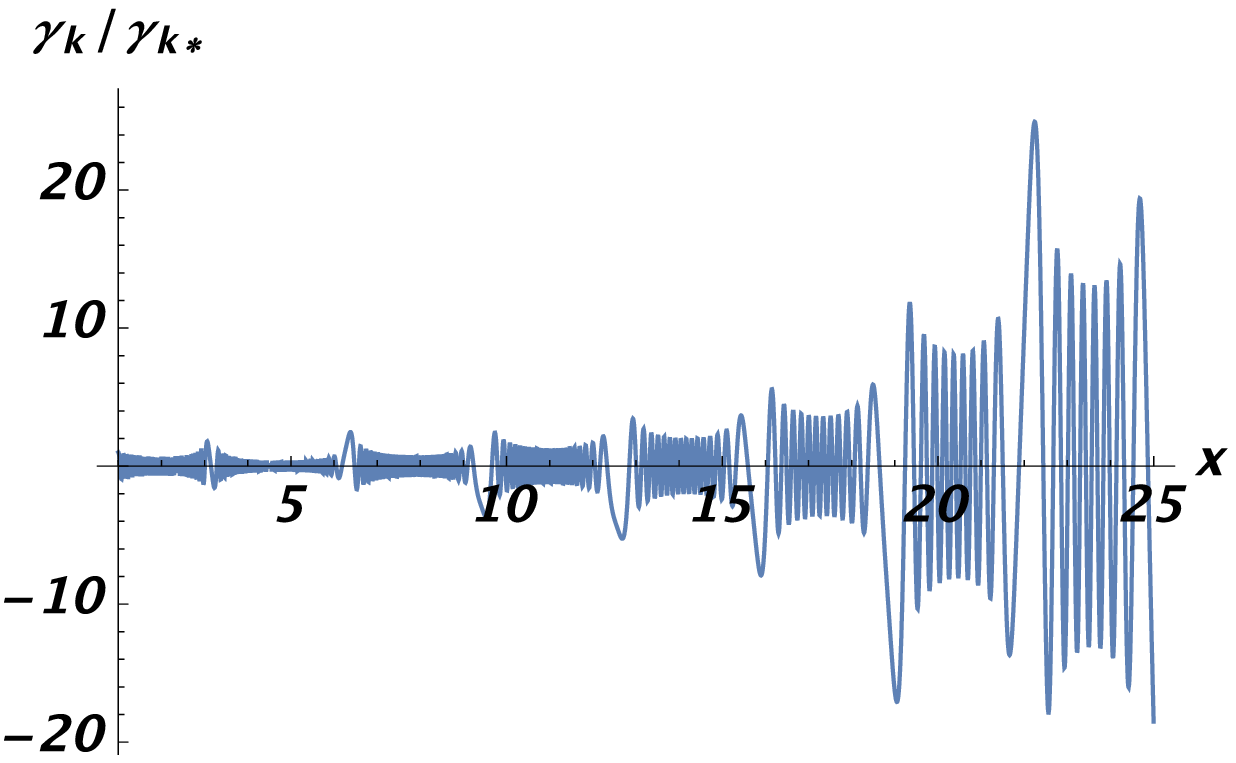}\includegraphics[width=0.25\textwidth]{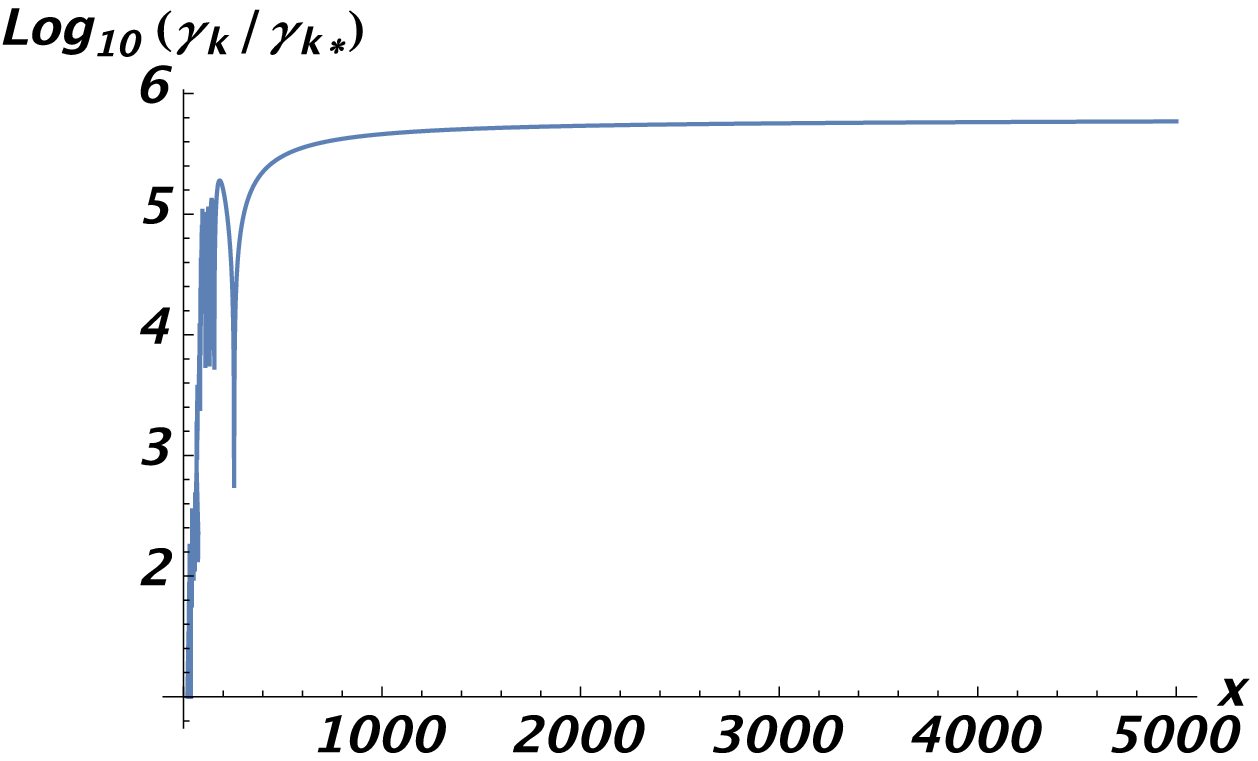}
\end{center}
\caption{The resonant amplification of tensor modes
during reheating. The horizontal axis is $x\equiv Mt$ and 
the vertical axis is the relative amplitude of the tensor modes
 $\gamma_k/\gamma_{k*}$, where $\gamma_{k*}$ is the amplitude
at the end of inflation. The parameters are $\xi=10^6$ and $\Gamma=0.05$,
with the initial condition ${d\gamma_k}/{dx}|_{x=1}=0$.}
\label{resonance}
\end{figure}

\begin{figure}
\begin{center}
\includegraphics[width=0.25\textwidth]{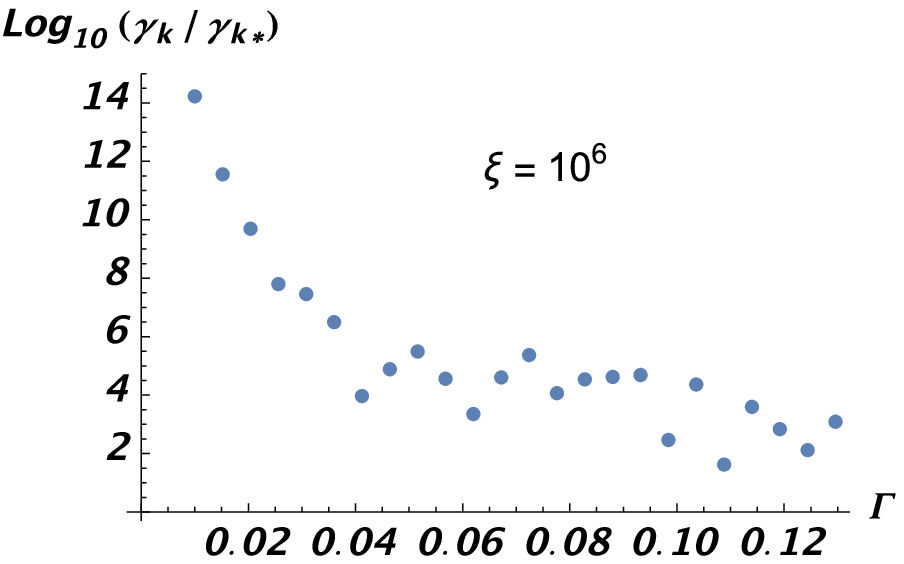}\includegraphics[width=0.25\textwidth]{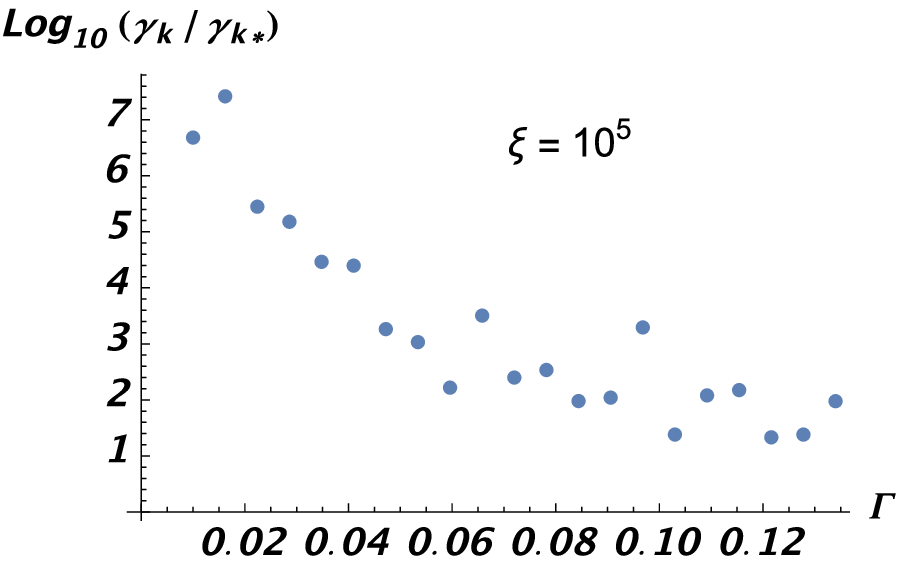}
\includegraphics[width=0.25\textwidth]{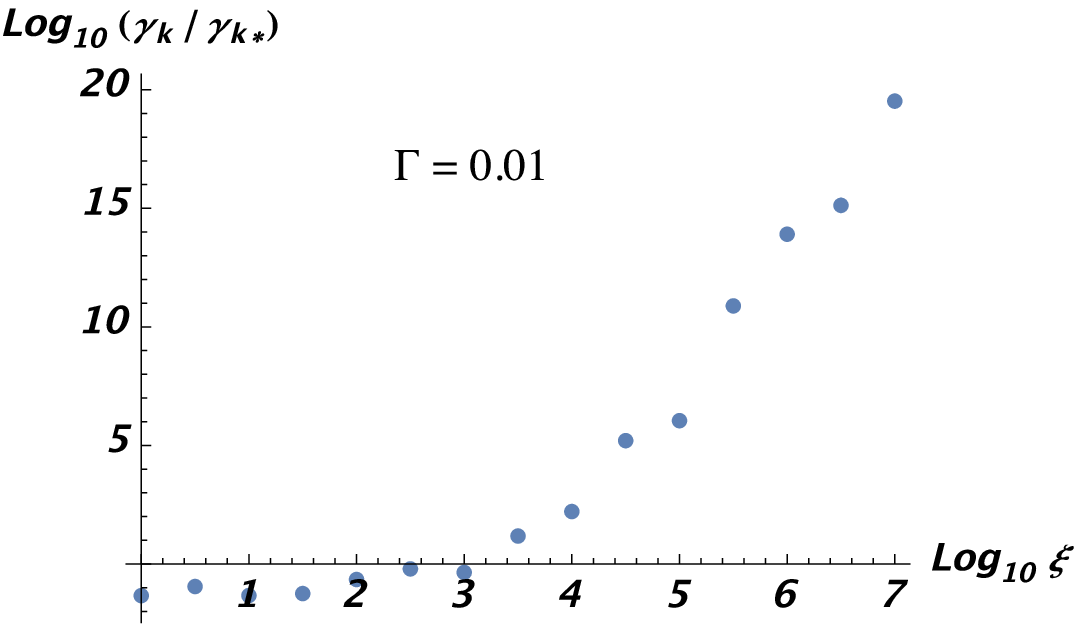}\includegraphics[width=0.25\textwidth]{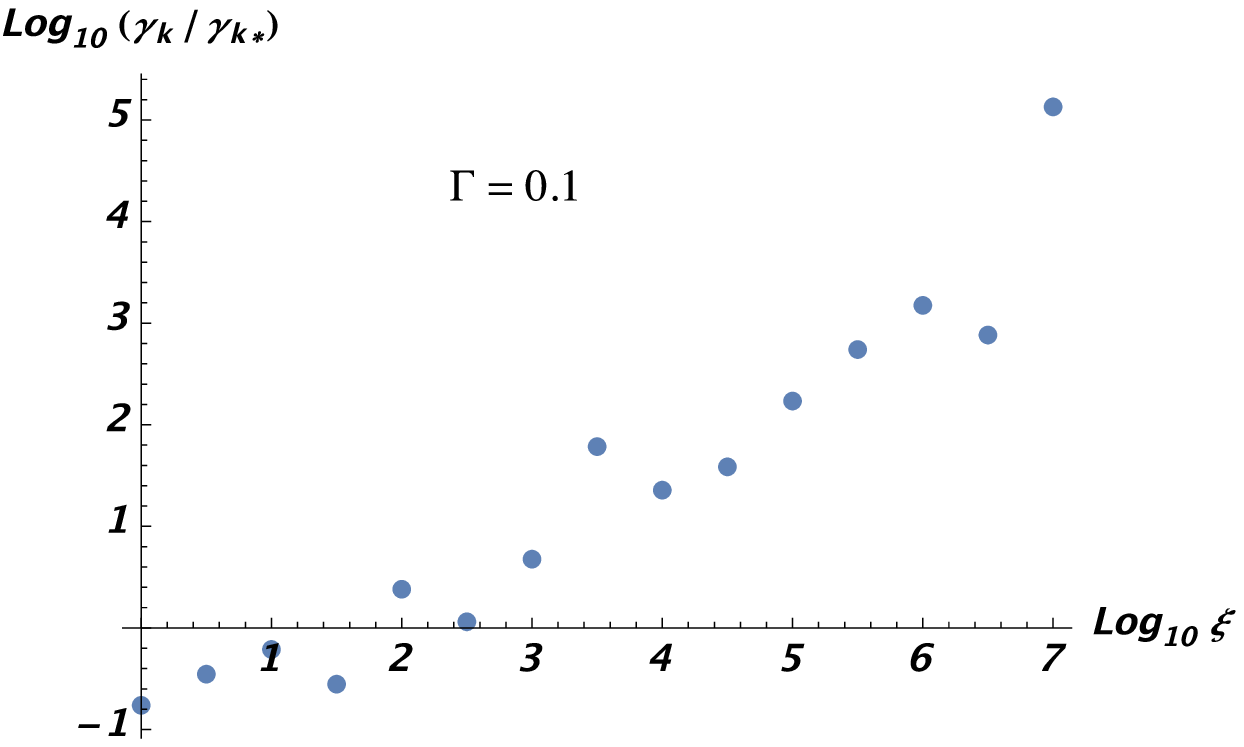}
\end{center}
\caption{The parameter dependence of the resonant amplification. 
 The horizontal axis is $\Gamma$ for the upper panels
and $\xi$ for the lower panels. 
The vertical axes is $Log_{10}(\gamma_k/\gamma_{k*})$ evaluated at $x=1000$.}
\label{decayrate}
\end{figure}

We have numerically solved Eq.~(\ref{mathieuGW}). The results are plotted in 
Figs.~\ref{resonance} and \ref{decayrate}.  
The resonant amplification factor depends on the value of $\xi$ and 
the decay rate $\Gamma$.  We can also read off the threshold 
for a significant resonant amplification is roughly $\xi>10^3$. 
The tensor modes stop growing in the large $Mt$ limit due to the decay of 
the inflaton. For those long wavelength modes whose gradient term is 
always negligible during reheating, the final power spectrum is still almost scale-invariant, 
as long as the graviton mass during inflation is small enough and 
thus the tensor tilt in Eq.~(\ref{ttilt}) is small.

Noted that the kinetic term of inflaton field  is canonical at leading order, which implies that in the massless limit $m_g^2\to0$, scalar and tensor just simply decouple. On the other hand, it has been previously proven that this theory smoothly reduces to GR 
in the massless limit due to the absence of vDVZ discontinuity \cite{Dubovsky:2004sg}. After reheating, $\phi\to0$, 
the graviton becomes massless and we recover GR.

{\bf Vector perturbation~} 
 To calculate the vector perturbation, we adopt the unitary gauge,
 in which the fluctuations of $SO(3)$  $St\ddot{u} ckelberg$ scalar fields are fixed to be 
zero, i.e. $\delta\varphi^i=0$. The quadratic action of the vector 
perturbation reads (in momentum space)
\begin{eqnarray}
S_V^{(2)}=\frac{M_p^2}{16}\int a^3k^2
\left[\dot{F_i}\dot{F_i}-m_g^2F_iF_i-\frac{4S_i\dot{F}_i}{a}
+\frac{4S_iS_i}{a^2}\right].
\end{eqnarray}
After integrating out $S_i$, we get
\begin{eqnarray}
S_V^{(2)}&=&-\frac{1}{16}M_p^2m_g^2\int a^3k^2F_iF_i .
\end{eqnarray}
This clearly shows that the kinetic term for vector perturbation was canceled out. It is by no mean of an accident, because the kinetic term of vector modes are prohibited by internal symmetry  $\varphi^i\to\varphi^i+\Xi^i\left(\varphi^0\right)$. 

{\bf Scalar perturbation~} 
In the scalar sector, $\alpha$, $\beta$ and $E$ are non-dynamical.
 After integrating them out, the quadratic action for the 
scalar perturbation in the uniform $\phi$ gauge (i.e. $\delta\phi=0$)
reads
\begin{eqnarray}
S_{s}^{(2)}=M_p^2\int a^3\epsilon
\left(\dot{\psi}^2-\frac{k^2}{a^2}\psi^2\right),
\end{eqnarray}
This is exactly the same as the one in GR with a single scalar field.
 In this gauge, $\psi$ is identical to the curvature perturbation on
the comoving slicing, ${\cal R}_c$, and the power spectrum is 
given by the same formula~\cite{Sasaki:1986hm,Mukhanov:1988jd},
\begin{eqnarray}
P_{\cal R}=\frac{H^2}{8\pi^2\epsilon M_p^2}.
\end{eqnarray}
Thus the tensor-to-scalar ratio we observe today is
\begin{eqnarray}
r= \frac{A\times P_{\gamma}}{P_{\mathcal{R}}}=16\epsilon\times A,
\end{eqnarray}
where $A$ is the resonant amplification factor of the tensor modes 
during reheating. For instance, with the parameters choice in 
Fig.~\ref{resonance}, the factor $A$ could be of the order of $10^{11}$. 
The variation of the inflaton per $e$-fold is 
\begin{eqnarray}
\frac{d\phi}{dN}=\frac{\dot{\phi}}{H}=\pm\sqrt{\frac{r}{8A}}.
\end{eqnarray}
Thus during 60 $e$-folds, $\phi$ traverses a distance 
$\Delta\phi\simeq15M_p\sqrt{2r/A}$. Hence a sizeable 
tensor-to-scalar ratio is possible even for a sub-Planckian excursion. 
 We conclude that the Lyth bound can be explictly evaded.

{\bf Conclusion and discussion~} 
In this letter, we have minimally extended GR to a theory
with a non-vanishing graviton mass term and proposed a mechanism to
enhance the primordial tensor perturbation from inflation in
the the early universe.  In our model, only the tensor perturbation
is affected, while the scalar and vector perturbations 
remain the same as the ones in GR. 
The graviton mass is assumed to be proportional to the inflaton during 
reheating, and hence its coherent oscillations give rise to a significant 
resonant amplification for all long wavelength modes on super-horizon scales.
Then we have numerically studied the dependence of the amplification factor 
on the graviton mass and the decay rate of inflaton during reheating.
We find that the Lyth bound can be explicitly evaded in our model. 

Our model contains three non-dynamical spacelike $St\ddot{u} ckelberg$ 
fields $\varphi^i$, which may formally become dynamical  if we include 
higher order derivative terms. 
During inflation, however,  the would-be new 
 degrees of freedom are supermassive and exponentially decay away. Thus we can safely integrate out
 these modes at low energy scale. On the other hand, to screen these would-be new degrees at late time, if necessary, we can simply add a tiny but non-zero constant to
 the mass term in Eq.~(\ref{mass}).
 The current upper bound of the graviton mass $m_g$ is 
about $10^{-20}$eV, from observation of Hulse-Taylor binary pulsar, 
PSR B 1913+16 \cite{Sutton:2001yj}. 

At the nonlinear perturbation level, we expect that the graviton mass term 
will introduce several new interaction terms.
 It will be interesting to study its possible imprints in, e.g. the
 non-Gaussianity of CMB anisotropies. 
As for models beyond the minimal model, a massive graviton generically
induces non-trivial scalar and vector perturbations.
We plan to study such possibilities and their possible observational effects 
in future.


\begin{acknowledgments}
{\bf acknowledgments}
We would like to thank A. De Felice, X. Gao, S. Mukohyama, 
T. Tanaka, A. Taruya, and N. Tsamis for useful discussions.
\end{acknowledgments}

\end{document}